# Study on the ERP Implementation Methodologies on SAP, Oracle NetSuite, and Microsoft Dynamics 365: A Review

Madabattula Archana

Global Doctor of Business & Administration

Swiss School of Business and Management, Switzerland

madabattula@ssbm.ch

Prof Dr VijayaKumar Varadarajan,

Swiss School of Business and Management, Switzerland

vijayakumar@ssbm.ch

Sai Sravan Medicherla

Dept. of Information & Communication Technology

Manipal Institute of Technology

Manipal, India

sai.medicherla@learner.manipal.edu

## ABSTRACT

There are Top three vendors in the ERP market E.g., SAP, Oracle Net Suite and Microsoft dynamics 365 leading the Global ERP market. [1] While analyzing the ERP selection and implementation trends, it is critical that any organization looking to implement an ERP system assesses the vendors through the lens of its own organization's specific requirements. When choosing the right ERP, few things must be taken into consideration like the Time Budget and resources. The research paper analyses each phase and compares the methodologies of SAP, Oracle Net Suite and Microsoft Dynamics 365 and suggests the best methodologies to be practiced for any ERP projects. Like a poorly planned trip, if you don't have effective methodology, you can expect a negative impact on your implementation, solution quality, and business satisfaction. Wrong choice of methodology may lead to poor decision-making, best practices may not be followed, and teams may be disjointed in the implementation, which can cause delays. Choosing the right ERP methodology is the key. Methodology is the lifeline for successful project implementation.

## INTRODUCTION

With the growing and expansion of the business in the global market, business need a faster application to monitor and to enhance their decision-making process and to Invest in Enterprise Resource Planning. **What is ERP [3]:** ERP means Enterprise Resource Planning. As the name itself suggest it's planning the Enterprise Resources. Enterprise's means Company or an Organization, and it has many departments like Accounting and Finance, Sales and Marketing, Human

> METHODOLOGY IS THE LIFELINE FOR A PROJECT IMPLEMENTATION.

Resources and Manufacturing department etc. As company expand globally and business is open for International customers then it would be difficult for companies to manage the business and this is when ERP application makes is way to a automate business processes and provide internal control for all internal stakeholders and departments , where the central database that collects all the business data will help the Management to make the right decision .The right ERP application for the business will help the client to grow faster, cost effectiveness and high productivity.

ERP application can help organizations bring together different activities under a single platform, An ERP system binds together all these processes with a unified interface. This makes it easy for users to access data with a centralized dashboard and features like access control and increased data security.

**Why methodology is important?** 50% of the ERP implementations fail due to choice of the ERP Methodologies.[2]

**Define Methodology** -

The term "project management methodology" was first defined in 1960s, when various business organizations have begun to seek effective ways that could simplify the realization of benefits out of businesses in a structured and unique entity (which was later called "Project"). Methodology is the core of project execution that allows the team to take the right path towards implementation. A methodology is a prescriptive definition of series of activities to be undertaken for a given project.

# METHODOLOGIES FOR ERP IMPLEMENTATION

It describes the use of a collection of methods to achieve predictable outcomes. The global Enterprise Resource Planning market is valued $45.91 billion in 2022 and is projected to reach $59.97 billion by 2027 according to https://www.statista.com.

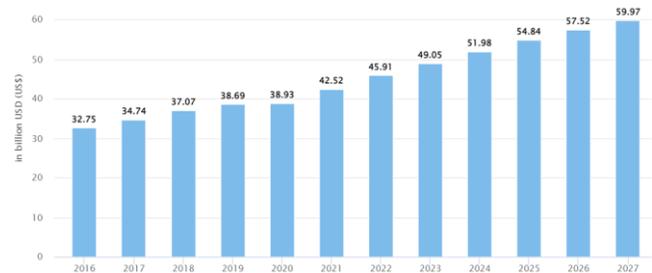

The Word's Top Leading ERP's are SAP having market share of 19.3%, global Oracle NetSuite with 11.2%, Microsoft Dynamics 365 market share is around 6.2%. According to https://www.statista.com

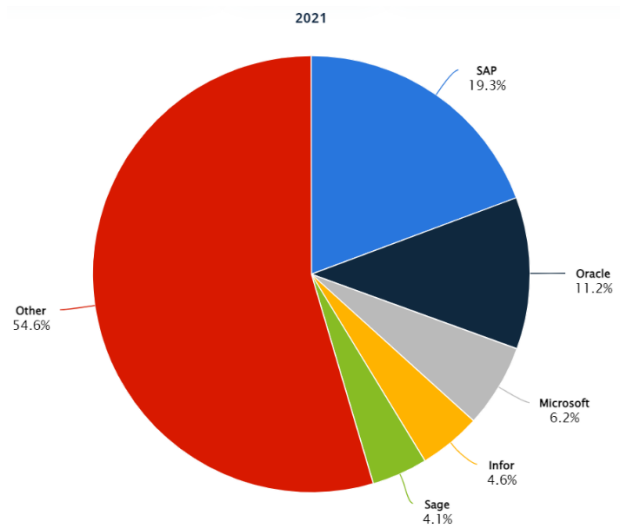

According to [1] SAP is a software business provider that offers software applications and services that helped organizations in more than 25 industries to run their businesses in a successful approach.

---

**RESEARCH OUTLINE**
My research paper outlines on the methodologies used to implement the ERP application and the focus is more on the process, the base of each of the methodologies used by SAP, Oracle NetSuite, Microsoft Dynamics 365

---

## SAP:

SAP software applications and services are used among more than 47,800 customers in 120

different countries and SAP Company is listed on several stock exchanges.

SAP has methodologies to implement in the application for their business.

SAP Methodologies:

**Phase 1 Project Preparation** – We make the necessary preparation before we met the client
- We obtain the support from the senior-level management/stakeholder support.
- Identifying clear project objectives
- Architect an efficient decision-making process
- Creating an environment suitable for change and re-engineering
- Building a qualified and capable project team.

**Phase 2 Business Blueprint -** We meet the Client in this Stage.[5]
This is the external meeting and the this is the first meeting with the client.
This phase is to help extract pertinent information about company that is necessary for implementation. Here, we will gather all the information about the how the business runs. These blueprints are in the form of questionnaires that are designed to probe for information that uncovers how your company does business. As such, they also serve to document the implementation.
- Each business blueprint document essentially outlines your future business processes and business requirements. The kinds of questions asked are germane to the particular business function.
- The questions are retrieved from the AS – IS document, which are unambiguous, which needed more details etc.

**Phase 3 Realization** – In the Realization phase,
With the completion of the business in phase 2, "functional" experts are now ready to begin configuring SAP. The Realization phase is broken in to two parts.

Your SAP consulting team helps you configure your baseline system, called **the baseline configuration.** Your implementation project team fine-tunes that system to meet all your business and process requirements as part of **the fine-tuning configuration.**

The initial configuration completed during the base line configuration is based on the information that you provided in your blueprint document. The remaining approximately 20% of your configuration that was not tackled during the baseline configuration is completed during the fine-tuning configuration.

Fine tuning usually deals with the exceptions that are

| |
|---|
| Phase 1 Project Preparation |
| Phase 2 Business Blueprint AS IS – TO BE |
| Phase 3 Realization |
| Phase 4 Final Preparation |
| Phase 5 Go-Live and support |

not covered in baseline configuration. This final bit of tweaking represents the work necessary to fit your special needs.

**Configuration Testing-**
With the help of your SAP consulting team, you segregate your business processes into cycles of related business flows. The cycles serve as independent units that enable you to test specific parts of the business process. You can also work through configuring the SAP implementation guide (IMG). A tool used to assist you in configuring your SAP system in a step by step manner.

**Knowledge Transfer-**
As the configuration phase ends, it becomes necessary for the Project team to be self-sufficient in their knowledge of the configuration of your SAP system.

Knowledge transfer to the configuration team tasked with system maintenance (that is, maintenance of the business processes after Go-live) needs to be completed at this time. In addition, the end users tasked with using the system for day-to-day business purposes must be trained.

**Phase 4 Final Preparation**

- As phase 3 merges into phase 4, you should find yourselves not only in the midst of SAP training, but also in the midst of rigorous functional and stress testing.
- Phase 4 also concentrates on the fine tuning of your configuration before Go-live and

more importantly, the migration of data from your old system or systems to SAP.
- Workload testing (including peak volume, daily load, and other forms of stress testing), and integration or functional testing are conducted to ensure the accuracy of your data and the stability of your SAP system. Because you should have begun testing back in phase 2, you do not have too far to go until Go-live.

Now is an important time to perform preventative maintenance checks to ensure optimal performance at your SAP system. At the conclusion of phase 4, take time to plan and document a Go-live strategy. Preparation for Go-live means preparing for your end-user's questions as they start actively working on the new SAP system.

**Phase 5 Go-Live and support –**

Go-Live activity will be smooth and breeze if all the previous phases have been executed under a strict monitoring process. If all the previous phases are not dealt with the way they were supposed to be then it is altogether a different ball game.

Introducing a new system or tool or service to entire different individuals, in this case, preparation is the key. Within this preparation, one has to make sure that they have answered all the What If scenarios within the business processes and how is the tool or system is capable of answering or supporting in those situations.

Further, processing maintenance contracts and documenting the overall Go-live process is vital in this phase.

.
# ORACLE NETSUITE METHODOLOGIES:

In the Methodologies for the first phase practiced at Oracle Net Suite is ENGAGE.

| Stage 1: Engage |
| Stage 2: Drive |
| Stage 3: Enable |
| Stage 4: Convert |

**Oracle NetSuite Stage 1: Engage**

The Objective of this Session is Engage with the customer
Set expectations and framework for a successful Implementation. The milestone Successfully set expectations and framework Trust is formed between Customer and Implementation Team which is focused more on the Customer Engagement. The Activities are:
- Staff Project
- Engage Account Manager
- Customer attend LCS Company Pass: Getting Started Webinar
- Reverse Sales to PS KT
- Conduct Kick-Off Meeting
- Complete Getting Started Session
- Install Suite Success Bundles
- Finalize Project Plan
- Hold User Adoption Discussion
- Develop Data Migration Strategy
- Begin Data Migration
- Confirm project team has begun Industry Fundamentals
- Conduct Education and Adoption Strategy Meeting
- Conduct Personalization Sessions
- Build Awareness
- Complete Engage Checkpoint
- Develop Functional Design Document (if appropriate)

In the Oracle Net Suite, Reverse **Sales to CS KT** (Customer Success) (Knowledge Transfer) session. – in this session the Delivery team meets the Sales team. This is an Internal meeting. In this meeting the Delivery Team review the License owned, customer website, Sales artifacts etc. The Objective of this meeting is to know about the client personality, why they choose Net Suite and to know about the pain points of their business.

Then Kick – off meeting is conducted which is the first one to one meeting with the client where we confirm the customers goals and initiatives. we review the scope of the project. The Kickoff meeting is the official start of project delivery and marked as Day 1 of the project. Establish the project methodology and governance followed by Bundle installation session and immediately Getting started session starts, Allow the customer to log into their NetSuite account for the first time. • Grant access to the NetSuite CS team. • Highlight key NetSuite features. • Introduce the Data Migration templates. • Emphasize that all customer data is stored in the File Cabinet. • Introduce Support, SuiteAnswers, and Suite

Ideas. Through the Engage Phase, Trust is formed between the Delivery team and the Customer.

**Oracle NetSuite Stage 2: Drive**

In the Methodologies for the second phase practiced at Oracle Net Suite is DRIVE, The Objective of this session is to Lead customer through configuration requirements and initial walkthrough of NetSuite. The milestone Customer understands configuration requirements and basics of NetSuite

Lead customer through configuration requirements and initial walkthrough of NetSuite

- Complete Configurations
- Develop Scripts (if appropriate)
- Continue Data Migration
- Conduct Process Walkthrough(s)
- Develop Training Approach and Materials
- Conduct Learning Action Planning Meeting
- Develop User Acceptance Plan
- Confirm project team has begun Administrator Fundamentals
- Drive Understanding
- Complete Drive Checkpoint

During the Personalization session, this is business requirement gathering session where the Delivery team gathers all the business requirement from the Process owners to configure the business based on their requirements in the Oracle Net Suite. The Delivery team will lead the session and they will ask questions that have the dispute in the blueprint, they will ask questions that were not clear to them in the blueprint. They will ask questions to the customer that were unambiguous to them in the blueprint. Once the requirement is gathered the Delivery team will configure the business based on the customer requirement and will demonstrate the business in the NetSuite in the Process walkthrough Session. This Process Walkthrough session will make the client understand how his business is going to be in NetSuite environment. After successful completion of the Process Walkthrough we guide the client for UAT.

**Oracle NetSuite Stage 3: Enable**

In the Methodologies for the third phase practiced at Oracle Net Suite is ENABLE, prepare customer for Go-Live, Train end users, Support customer UAT. The milestone is Completing the UAT and the End user Training. The Activities are:

Continue Data Migration
Conduct Final Process Walkthrough
Conduct Education Services KT
Provide User Acceptance Testing (UAT) Guidance
Basic Usability Training for UAT
Conduct Pre-UAT Training (optional)
Conduct UAT
User Training
Provide Support Video
Empower End Users
Complete Enable Checkpoint

In the Enable Stage, The Process Walkthrough 2 is conducted. This session is between the customer and the Delivery team and in this session the Customer will lead the session, which meant that the customer would show the Process of their business to the Delivery team. The objective of this session is to make sure that the customer familiar's themselves with the Net Suite application and also to make sure that they have understood the NetSuite application.

**Oracle NetSuite Stage 4: Convert**

In the Methodologies for the last and fourth phase practiced at Oracle Net Suite is Convert, the objective of this Convert stage is to Confirm business readiness and instill confidence and transfer system ownership to Customer. The Activities are:
Complete Data Migration
Confirm any Additional Implementation Readiness Training
Complete Cutover Checklist
Conduct Cutover
Provide Post Go-Live Support
Enter Info in "As Built"
Customer Transition Meeting
Harness Ownership
Post Go-Live Follow Up Training Sessions
Complete Convert Checkpoint
Transition to NS Support and/or ACS

# MICROSOFT DYNAMICS 365 METHODOLOGIES

Success by Design maps the Dynamic 365 implementation lifecycle into four methodology phases:

| |
|---|
| Stage 1: Initiate |
| Stage 2: Implement |
| Stage 3: Prepare |
| Stage 4: Operate |

In this section and the following sections, we outline the Success by Design phases, their relationship to Success by Design reviews, and the desired outputs and outcomes.[6]

Core implementation team: This is the team doing the actual execution of the project. For any Dynamics 365 implementation project, core implementation teams should include project manager(s), business subject matter experts, solution architects, functional and technical consultants, developers, testers, change management advisors, and user champions.

**Microsoft Dynamics 365 Get ready to start Stage 1: Initiate**

In the Initiate phase,
- Kick off
- Requirement analysis
- Fit gap analysis
- Customer kick off
- The project team will be gathering and validating business requirements
- finalizing the high-level solution approach, making inroads to define all in-scope workstreams, and updating the project plan to reflect these updates.
- When the project team has produced the high-level solution design and the related project workstreams are more or less defined, Success by Design begins with the Solution Blueprint Review.

- **Microsoft Dynamics 365 Design and Build Phase**
  **Stage 2: Implement**

In the Implement phase, we perform the following:
- Code
- Configuration
- Data modeling
- Solution Performance
- Integration
- the project team is focused on building the solution per the agreed-upon solution design and scope.
- Implementation Reviews are introduced in this phase, having been informed by the findings and recommendations of the Solution Blueprint Review.
- Implementation Reviews are used to more deeply address questions related to the specific aspects of the solution design (data model, security, integration) and implementation practices (ALM, testing strategy). Implementation Reviews are meant to fully address the risks identified during or after the Solution Blueprint Review but before the solution build is too far along.

**Microsoft Dynamics 365 Deploy Stage 3: Prepare**

In the Prepare phase,
- UAT
- Mock go live
- Cutover
- Testing and Acceptance
- Go live planning
- User readiness
- Cutover planning

The solution has been built and tested and the project team is preparing for the final round of user acceptance testing (UAT) and training.
- Additionally, all necessary customer approvals have been granted, information security reviews completed, the cutover plan defined (including go/no-go criteria), mock go-lives scheduled, the support model ready, and the deployment runbook completed with tasks, owners, durations, and dependencies defined.
- At this point, the project team uses the Success by Design Go-live Readiness Review to identify any remaining gaps or issues.

The Prepare Stage of the Microsoft Dynamics 365, Implementation Reviews are meant to fully address the risks identified during or after the Solution Blueprint Review. This phase practiced by Microsoft Dynamics 365 analyses the potential risks. This is one of the best phases in the Success by Design phases.

**Microsoft Dynamics 365 Live Phase Stage 4: Operate**

In the Operate phase,
- the customer solution is live.
- Service the solution
- Support
- Bug tracking
- Solution health
- Usage
- Maintenance
- Post Go-Live readiness review

CONCLUSION Success by Design equips project teams with a model for technical and project governance that invites questions and reflection, which leads to critical understanding of risks that might otherwise go unnoticed until too late in the project. Considering the pace of cloud solutions and the investment that organizations make in Dynamics 365 software and implementation costs, even the most experienced customers and partners with the best methodologies will benefit by incorporating Success by Design into their projects.

To achieve that, business applications must do more than just separately run your back office, marketing, supply chain, or even field operations. They must give you the agility to remove every barrier in your way. When this happens, business applications become more than just operational solutions. They become resilient solutions that enable the digital feedback loop and adapt to customer demands, delivering stronger, more engaging experiences around the products you make and services you provide. In Microsoft Dynamics 365, the Success by Design phases and their relationship to Success by Design reviews the desired outputs and outcomes with a user-friendly approach.

**Comparison between the ERP Implementation Methodologies of SAP, Oracle NetSuite and Microsoft Dynamics 365- My research conclusion:**

According to me, Knowledge Transfer should start from the Sales Team.
When compared to all the other Methodologies used by SAP and Microsoft Dynamics 365, Reverse to Customer Success Knowledge Transfer, is the best practice used by Oracle NetSuite because here the Delivery team is well-prepared prior meeting the client. Since the Sales team is also involved its gives them a higher overview of the customer expectations. In the other Methodologies like SAP and Microsoft Dynamics 365, the Sales team is not involved.

        This Methodology followed by only Oracle NetSuite, give a lot of importance to the Customer Engagement. This session is pre preparation before meeting the client in person. Since the Sales team is also involved its gives them a higher overview of the customer expectations. By preparing the Delivery Team for the next meeting with the client it sets the preparedness and strong effectiveness to know the customer better , their personalities and the pain points in the business and why they choose this ERP , and set expectations and framework for a successful implementation. This is the foundation for any ERP methodology.

All the ERP Vendors, follow the Project gathering phase as the initial  but When compared to all the other Methodologies used by SAP and Microsoft Dynamics and Oracle NetSuite ,one of the best practice used  here is the Revere Sales to PS KT session.

In the Enable Stage of Oracle Net Suite, The Process Walkthrough 2 is conducted. This session is between the customer and the Delivery team and in this session the Customer will lead and where the customer will showcase the business process in Net Suite to the delivery team. This is one of the best practices as it will let us know how well the client has understood the application and how user friendly the client is with the Suite.  This is such an amazing session which is only followed by the Oracle NetSuite methodologies which is not followed by other ERP's like Microsoft Dynamics 365 and SAP. One of the critical success factors for a project is also training. Through the process walkthrough 2 and Reverse to Sales Knowledge Transfer, many failed implementations can be reduced.

        I suggest that any ERP Methodology should have Reverse to Sales KT and Process Walkthrough 2. For Oracle NetSuite, Customer Engagement lead to Customer Success.  For Microsoft

Dynamics 365, every business is of creating great customer experiences. In Oracle NetSuite, Following the best practices suggested by the Suite Success for the customer's business combined with Customer Engagement brings Customer success.[7]

In SAP, Process and Sustainability is the Economics to bring the process and application technology integration of all the departments like Sales and Marketing, Finance, Production planning and so on.

All the three ERP Vendors are pioneer in their own way and the methodology used determines the success of the project implementation.